\documentclass[twocolumn]{aa}			
\usepackage{graphicx}

\usepackage{txfonts}
\usepackage{natbib}
\bibpunct{(}{)}{;}{a}{}{,}
\def\cmmd{\rm {cm^{-3}}}
\def\cmmt{\rm {cm^{-2}}}

\def\s-1{\rm {s^{-1}}}

\def\kms {\hbox{${\rm km\,s}^{-1}$}}

\def\asec{$''$}
\def\pow#1#2{#1$\times$10$^{#2}$}

\begin{document}

   \title{High-resolution HNC 3--2 SMA observations of Arp~220}


   \author{S. Aalto\inst{1}
          \and
          D. Wilner\inst{2}
	  \and
	  M. Spaans\inst{4}
	  \and
          M. C. Wiedner\inst{3}
	  \and
	  K. Sakamoto\inst{2,5,6}
	  \and
	  J. H. Black\inst{1}
	  \and
	  M. Caldas\inst{1}
          }

   \offprints{S. Aalto}

   \institute{Onsala Space Observatory,
              S-439 94 Onsala, Sweden\\
              \email{saalto@chalmers.se}
        \and
        Harvard-Smithsonian Center for Astrophysics, 60
Garden Street, Cambridge, MA 02138, USA
	\and
 I. Physikalisches Institut, Universität zu K\"oln, Z\"ulpicher Str. 77, D - 50937 K\"oln, Germany
         \and
	  Kapteyn Astronomical Institute, PO Box 800, 9700 AV Groningen, The Netherlands
	  \and
Institute of Astronomy and Astrophysics, Academia Sinica,
    P.O. Box 23-141, Taipei 10617, Taiwan
\and
National Astronomical Observatory of Japan, Mitaka, Tokyo 181-8588, Japan
	  }
   \date{received, accepted}


  \abstract
  {}
   {We study the properties of the nuclear molecular gas of the ultra luminous merger
   Arp~220 and effects of the nuclear source on gas excitation and chemistry. Specifically,
   our aim is to investigate the spatial location of the luminous HNC 3--2 line emission
   and address the underlying cause of its unusual brightness.}
   {We present high resolution observations of HNC $J$=3--2 with the SubMillimeter Array (SMA).}
   {We find luminous HNC 3--2 line emission in the western part of Arp~220, centered on the
   western nucleus, while
   the eastern side of the merger shows relatively faint emission.
   A bright (36 K at $0.''4$ resolution), narrow (60 \kms)
   emission feature emerges from the western nucleus, superposed on a broader spectral
   component. A possible explanation is weak maser emission through line-of-sight amplification
   of the background continuum source.
There is also a more extended HNC 3--2 emission feature north and south of the
nucleus. This feature resembles the bipolar OH maser
morphology around the western nucleus. Substantial HNC abundances are required to explain the bright line emission
from this warm environment -- even when the high gas column density towards the
western nucleus is taken into account.
We discuss this briefly in the context of an X-ray affected chemistry and radiative excitation. 
}
   {The luminous and possibly amplified HNC emission of
   the western nucleus of the Arp~220 merger reflects the unusual, and 
   perhaps transient environment of the starburst/AGN activity there.
   The faint HNC line emission towards Arp~220-east reveals a real
   difference in physical conditions between 
   the two merger nuclei.
}

\titlerunning{HNC in Arp~220}
\maketitle
%

\section{Introduction}
\label{s:intro}

The closest of the ultra luminous galaxies (ULIRGs), Arp~220
($D$=73 Mpc, 1\asec=354 pc) provides
an excellent opportunity to study the properties of the interstellar medium
in the extreme environment at the heart of an ongoing galaxy merger.
High-resolution CO studies \citep[e.g.][]{sakamoto99} find that the molecular
gas is distributed in two counter-rotating molecular gas discs with radii of 100 pc, each
surrounding the two nuclei.
These discs are embedded in a larger ($\approx$ 1 kpc) gas disc which is
rotating around the dynamical centre of the system. The nature of the two
nuclei is different with the western one dominating at
mid-IR wavelengths, suggesting a compact starburst (and/or buried AGN)
and the eastern one less luminous, but
with deep silicate absorption features \citep[e.g.][]{soifer99}.  
\citet{de07} find a compact, hot dust source surrounding the
nucleus of Arp~220 west. They claim that this dust source surrounds an AGN and
that a considerable fraction of the total IR luminosity of Arp~220 stems
from this AGN.

Recently, it has been discovered that the HNC 3--2 line of Arp~220 is unusually luminous
compared to HCN 3--2 \citep{cerni06, aalto07a}. 
Bright HNC 3--2 line emission can be explained in several ways, alone or in combination:\\
a) The presence of significant quantities of cold ($T<30$ K) molecular gas \citep[e.g.][]{hirota98},
b) very large optical depths due to large columns of gas, c) an effect of mid-IR
pumping of HNC enhancing the luminosity, or d) an abundance enhancement
of HNC in an X-ray dominated chemistry near an AGN \citep[e.g.][]{aalto07a,
2005A&A...436..397M, 2006ApJ...650L.103M, 2007A&A...461..793M}.

Both HCN and HNC have degenerate bending modes in the IR. For HNC this
mode occurs at $\lambda$=21.5 $\mu$m with an energy level
$h\nu/k$=668 K and an $A$-coefficient of $A_{\rm IR}$=5.2 s$^{-1}$. For HCN the bending
mode occurs at $\lambda$=14 $\mu$m, energy level $h\nu/k$=1027 K
and $A_{\rm IR}$=1.7 s$^{-1}$. 
In a radiation environment where the intensity rises through the mid-infrared
toward a peak at longer wavelengths HNC undergoes absorption and fluoresence
at a much faster rate than HCN. As a result HNC can be pumped
more easily than HCN \citep[e.g.][]{aalto07a}.
The X-ray irradiation of molecular gas leads to a so-called
X-ray dominated region (XDR) \citep[e.g.][]{1996ApJ...466..561M,1996A&A...306L..21L} 
similar to photon-dominated regions (PDRs) associated with bright UV sources
\citep{1985ApJ...291..722T}. The more energetic (1-100 keV)
X-ray photons penetrate large columns ($10^{22}-10^{24}$ $\cmmt$) of gas and
lead to a different ion-molecule chemistry.
Models of XDRs by \citet{2005A&A...436..397M} indicate that
the HNC/HCN column density ratio is elevated ($> 1$)
compared to PDRs (where the HNC/HCN abundance ratio may be 1) and quiescent cloud
regions for gas densities around $10^5$
$\cmmd$. Hence, ULIRGS that contain an AGN and possess high gas densities
are likely candidates for luminous HNC emission.

In order to study the mechanisms behind the large HNC luminosity of Arp~220, we
undertook high resolution HNC 3--2 observations with the SMA. In Sect.~\ref{s:obs}
we present the observations and results.
In Sect.~\ref{s:maser1} and ~\ref{s:maser2} the luminous nuclear HNC emission is
discussed in terms of possible amplified emission in a hot, dense environment.
The origin of a more extended HNC emission is discussed in Sect.~\ref{s:ext} and 
HNC abundances are briefly discussed in Sect.~\ref{s:abundances}.


\section{Observations and results}
\label{s:obs}

We have observed the HNC 3--2 line (rest frequency 271.981 GHz) in Arp~220 
on 2006 June 30 (project 2006-03-S057) with the SubMillimeter Array (SMA) on
Mauna Kea, Hawaii.
We assigned velocity 5486 \kms to channel 36 (out of 72) corresponding to
frequency 267.093310 GHz thus using the optical velocity definition 
($V_{\rm opt}/c$=$\Delta\nu/\nu_{\rm observed}$). The optical velocity is approximately
100 \kms larger than the velocities for the radio definition. 
The observations were carried out in the very extended
array where the baselines ranged from 67 to 507 meters. There were
seven antennas in the array and the atmospheric opacity was $\tau_{225}$= 0.15, with low wind, and a
reasonable phase stability for this configuration. System temperatures were 200-700 K (DSB)
(with strong elevation dependence).
The quasar J1635+381 was used to calibrate both Arp~220 and J1613+342. Fitting
a Gaussian to the J1613+342 visibilities we find an effective seeing of about $0.''15$.
The data were calibrated with the MIRIAD package and then transferred to FITS format
for further reduction with the AIPS package.
The data were deconvolved using the AIPS CLEAN task IMAGR and the data were binned to a
velocity resolution of 30 \kms. For natural weighting, the beam size is
$0.''48 \times 0.''32$, omitting antenna 3,
which had performance issues. Dropping antenna 3 makes
the longest baseline 458 meters. At this resolution, the flux to temperature scale is: 120 K/Jy 
so that 0.3 Jy corresponds to 36 K. We used the AIPS task IMLIN for continuum
subtraction in the image plane.

\begin{table}
\caption{\label{t:flux} Continuum and HNC 3--2 Line Results}
\begin{tabular}{lll}
 & East nucleus & West nucleus\\
\hline
\hline \\ 
Continuum$^a$: & &\\
\, \, position (J2000) & $\alpha$:  15:34:57.31  & $\alpha$: 15:34:57.23 \\
                           & $\delta$: +23:30:11.4 &  $\delta$: +23:30:11.5 \\
\\
\, \, flux (mJy) & $57.6 \pm 16$ & $152 \pm 19$ \\
\\
Line$^b$: & &\\
\, \, peak flux & & \\
\, \, (mJy beam$^{-1}$)& 75 $\pm$ 50 & 330 $\pm$ 50\\ 
\\
\, \, integrated flux & & \\
\, \, (Jy \kms) & $18 \pm 9$ & $130 \pm 20$ \\
\hline \\
\end{tabular} 

a) Gaussians were fitted to the continuum images and the
eastern source was found to be slightly resolved while the western source was
unresolved. Removing a 10\% contribution from free-free and synchrotron continuum, we find
the corresponding beam-smeared brightness temperature of the dust emission
to be 16 K - similar to the 18 K
\citet{de07}  find for their 1.3 mm continuum source at somewhat higher resolution ($0.''3$). \\
b) The integrated flux in the naturally weighted map is 250 $\pm$ 80 Jy \kms. Out of these, 130 Jy \kms 
is associated with a 0.5\asec structure.

\end{table}

\subsection{Continuum emission}

Continuum emission at 277~GHz (upper side band -- the HNC 3-2 line was placed in the
lower side band) is detected towards both the western and eastern nucleus. 
The continuum results are presented in tab.~\ref{t:flux}.
These SMA observations were not aimed at obtaining absolute astrometry
and thus the positional uncertainty for the continuum peaks is $\approx 0.''2$.
We refer to \citet{sakamoto08} for more accurate positions of the continuum and
for a detailed discussion of the nature of the continuum sources. 

\subsection{HNC $J$=3--2 line emission}

\begin{figure}
\resizebox{7cm}{!}{\includegraphics[angle=0]{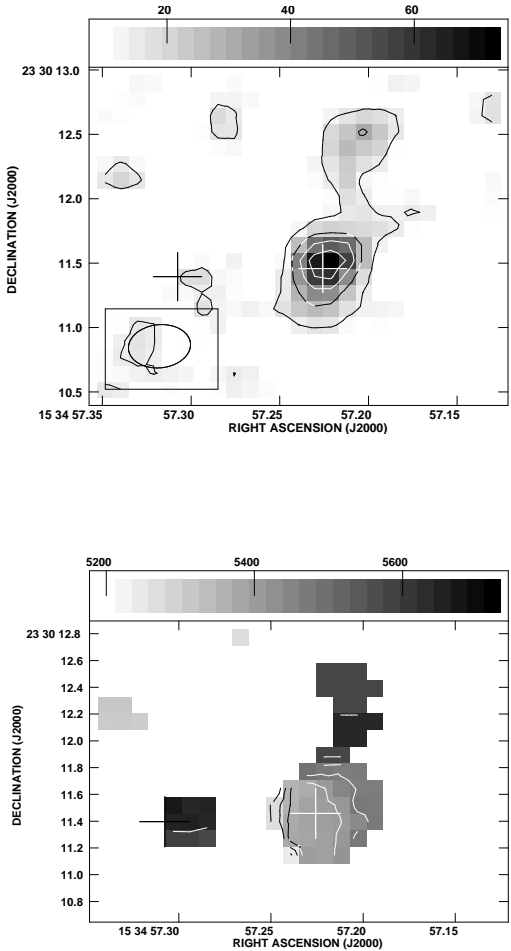}}
\caption{\label{fig:hnc32_na} Top panel: Integrated HNC 3--2 line emission (naturally weighted).
The grey scale ranges from 10 to 73 Jy beam$^{-1}$ \kms. Contour levels are 
7.3 Jy beam$^{-1}$ \kms $\times$(2,4,6,8). Crosses mark the position of the 277 GHz
continuum peaks. The synthesized beam is indicated in the box in the lower left corner.
Lower panel: The HNC 3--2 velocity field. The grey scale ranges from 5200 to 5720 \kms and the
contours range from 5252 to 5616 \kms with steps of 52 \kms.}
\end{figure}

\begin{figure}
\resizebox{6cm}{!}{\includegraphics[angle=0]{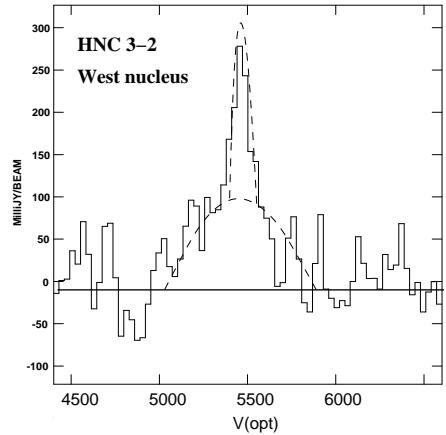}}
\caption{\label{fig:hnc32_spec} HNC $J$=3--2 spectrum of the western nucleus of Arp~220. 
The spectrum has been Hanning smoothed resulting in a reduced intensity in the narrow component. The spatial 
resolution is $0.''48 \times 0.''32$. The broad and narrow components are marked with a
dashed curve.}
\end{figure}

\subsubsection{Naturally weighted map}
\label{s:nat}

\begin{figure}
\resizebox{9cm}{!}{\includegraphics[angle=0]{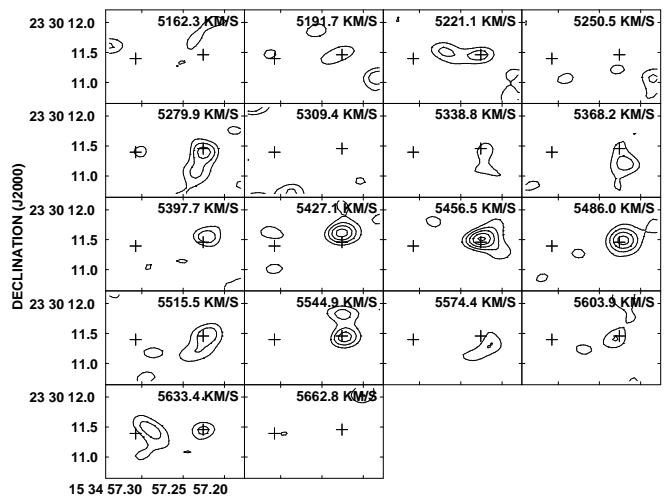}}
\caption{\label{fig:hnc32_na_ch} Naturally weighted channel maps,
not smoothed in velocity. Crosses
mark the radio continuum positions of the two nuclei. Contour levels are 
49 mJy $\times$ (2,3,4,5,6). Peak flux is (330 $\pm$ 50) mJy/beam.}
\end{figure}

In the {\it integrated line map} (fig.~\ref{fig:hnc32_na} upper panel), most of the HNC flux
is at the western nucleus of Arp 220 (see tab.~\ref{t:flux}).  The line peak
coincides with the western continuum peak.  The line peak
is about 250 mJy/beam in the Hanning-smoothed spectrum (Fig.2),
and it  is 330 +/- 50 mJy/beam (39 +/-6 K) in the unsmoothed data 
(the 30 km/s channel maps in Fig.3), at 5456 \kms.
The {\it velocity field} (fig.~\ref{fig:hnc32_na} (lower panel)) shows east-west rotation
around the western nucleus consistent with what has been seen previously for CO 2--1 and 3--2
\citep[e.g.][]{sakamoto99,de07,sakamoto08} from an edge-on rotating torus or disc.
Higher velocities to the north of the western nucleus, and in the eastern nucleus, are
also consistent with previous CO results.
A {\it channel map} is shown in fig.~\ref{fig:hnc32_na_ch} where we can see that the brightest line
emission can be found in the velocity range 5400 \kms to 5550 \kms - with a peak at 5456 \kms.
In total, line emission can be found from
5200 to 5633 \kms. From the channel map we see that emission is present in the nucleus
itself, and also extends to the north and south.

\subsubsection{Tapered map}

\begin{figure}
\resizebox{6cm}{!}{\includegraphics[angle=0]{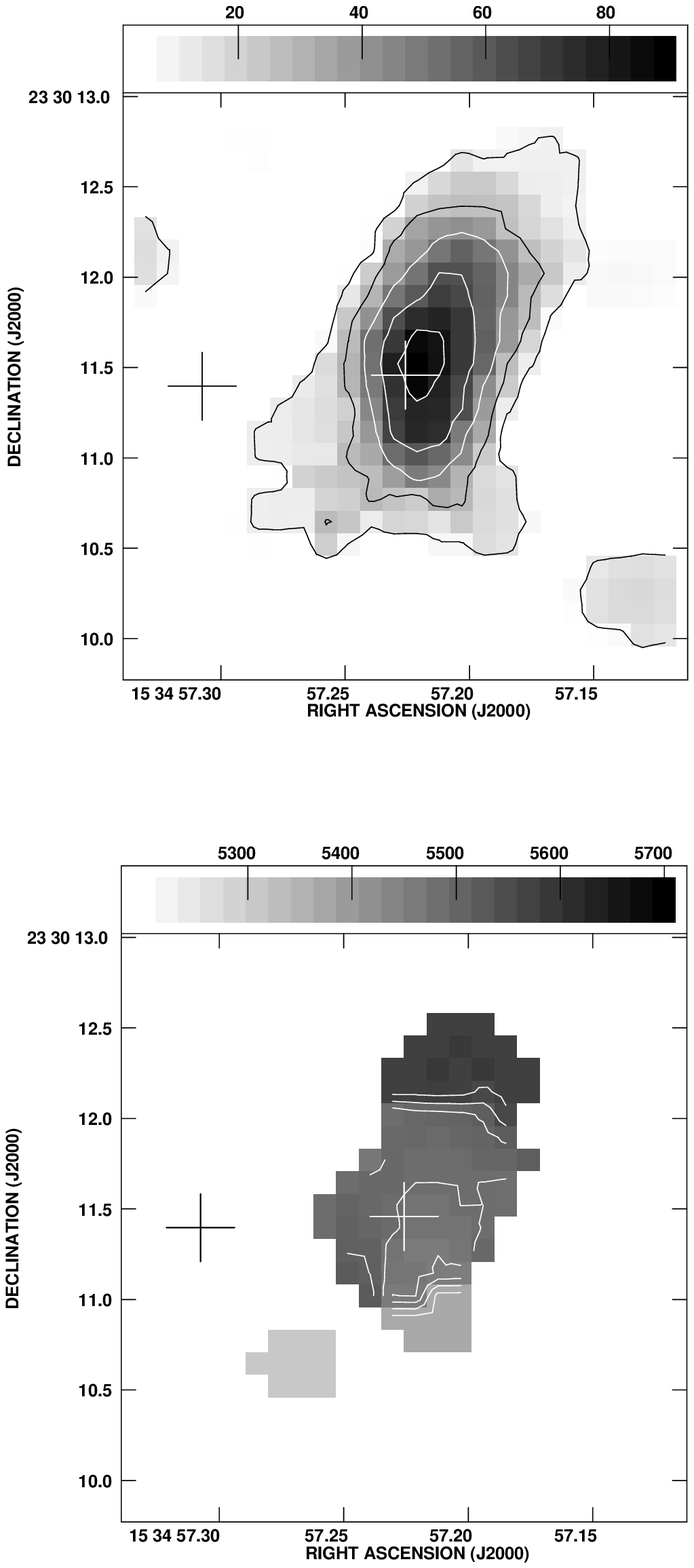}}
\caption{\label{fig:hnc32_tap} Top panel: Tapered integrated HNC 3--2 line emission
(beam $0.''77 \times 0.''67$ with PA=20$^{\circ}$).
The grey scale ranges from 5 to 90 Jy beam$^{-1}$ \kms. Contour levels are 
9.0 Jy beam$^{-1}$ \kms $\times$(1,3,5,7,9).  The rms noise in the map 10 Jy beam$^{-1}$ \kms making the
first contour 0.9$\sigma$ and the second 2.7$\sigma$. Crosses mark the position of the 277 GHz
continuum peaks.
Lower panel: Tapered velocity field. The grey scale ranges from 5200 to 5700 \kms and the
contours range from 5300 to 5490 \kms with steps of 27 \kms.}
\end{figure}

To study the more extended part of the emission we tapered the map
to a resolution of $0.''77 \times 0.''67$ where 0.3 Jy corresponds to
11 K (see fig~\ref{fig:hnc32_tap} top panel).
The integrated peak intensity is then shifted somewhat off the western nucleus
and the shape of the emission (and the velocity structure) appears dominated by a
north-south feature.

\subsection{Single dish spectrum and recovered flux}
\label{s:single}

\begin{figure}
\resizebox{8cm}{!}{\includegraphics[angle=0]{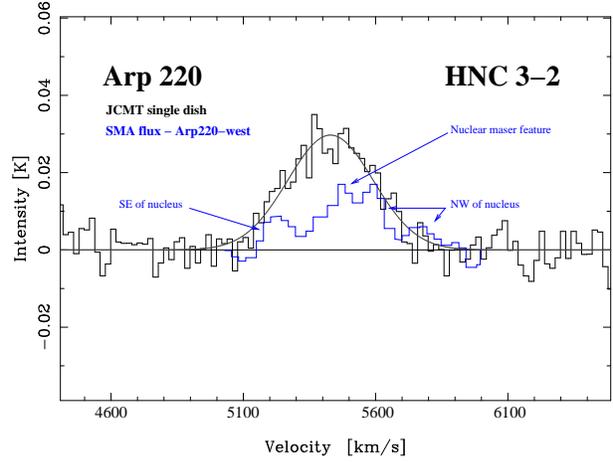}}
\caption{\label{fig:single} Top spectrum (in black):JCMT HNC $J$=3--2 spectrum of Arp~220,
the scale is in $T_{\rm A}^*$. The velocity resolution has been smoothed to 30 \kms and the velocity
scale is $V_{\rm opt}$. Lower spectrum: SMA naturally weighted spectrum summed over
the western nucleus and regions north and south of the western nucleus.}
\end{figure}

Single dish data on HNC 3--2  were reported in \citet{aalto07a} and obtained with the 
JCMT in April 2005. The integrated intensity translates to 351 Jy \kms
($S=27.4 T_{\rm A}^*$) where
the integrated intensity is 12.8 K \kms in the $T_{\rm A}^*$ scale.
In fig.~\ref{fig:single} we show the JCMT single-dish spectrum and the SMA (naturally weighted)
flux-summed spectrum for the western region of Arp~220.
Note the difference in line shape and that
there appears to be a larger fraction of blue-shifted emission missing in the naturally
weighted interferometer map.
The contribution from the eastern nucleus is not included in fig.~\ref{fig:single}. Its
contribution is small and would add 10\% flux to the higher velocity part of
the spectrum.

\section{Discussion}

The HNC 3--2 spectrum of the western nucleus shows a broad feature upon which
a narrow, luminous feature ($V_{\rm opt}$=5440 \kms) is superposed
(see Sect. ~\ref{s:nat} and fig.~\ref{fig:hnc32_spec}).
The peak brightness temperature of the HNC 3--2 emission feature is $\approx$39 K in
our $0.''4$ beam. 
In the same velocity range, CO 2--1 and 3--2 (at similar resolution) instead 
show deep, narrow (100 \kms) absorption \citep{de07,sakamoto08}.
The deepest CO 2--1 absorption occurs at 5450 \kms, but other, less prominent, 
absorption features show up at
higher velocities, In CO 3--2 the absorption occurs in the range
5400-5450 \kms (transferred from the $V_{\rm LSR}$ of \citet{sakamoto08}.
\citet{de07} suggest
that this is due partly to absorption of hot continuum, and CO self-absorption,
implying that the CO-West disc has a temperature gradient increasing inwards.

\par
{\it We suggest that the luminous, narrow HNC 3--2 emission features could be caused by
weak maser action in the cooler part of the nuclear disc - amplifying instead of absorbing
the background continuum from the inner hot, dusty part of the disc.} 
If the narrow,
luminous feature is indeed emerging only in front of the $0.''19$ continuum
source, the undiluted brightness temperature is 156 K. 

Previous claims of weak masering in other molecules (besides OH) include 
methanimine \citet{salter08}, 
and formaldehyde \citet{araya04}.

\subsection{A simple model}
\label{s:maser1}

\citet{de07}
propose a model of the western nucleus consisting of: An inner circum-nuclear disc (CND)
which is hot (dust temperature $T$=170 K) and with an average gas 
density $n$=\pow{1}{5} $\cmmd$ and H$_2$ column density $N({\rm H}_2)=10^{25}$ $\cmmt$.
Its rotational velocity is $V$=370 \kms and radius 30 pc.
The CND is surrounded by a cooler ($T$=50 K) outer disc or torus with average gas density 
$n$=\pow{1}{4} $\cmmd$, $N({\rm H}_2)=10^{24}$ $\cmmt$, radius 90 pc and has a rotational
velocity of $\approx$250 \kms. From now on we will refer to
these two features as the CND and the surrounding torus, ST. Thus in this scenario, the
CO absorption
occurs in the cooler ST obscuring the hot dusty nucleus. The division of the western nucleus into these
two features is somewhat misleading since to view them as two separate entities with constant temperature and
density is likely incorrect. However, for the purpose of a simple model the ST/CND model
will suffice and we adopt it for the maser discussion here. We have also assumed that the
rotation velocity of the ST falls linearly from 370 \kms at the radius of the CND
to 250 \kms (i.e. from 30 to 90 pc). The detailed dynamics of the western nucleus
is not known, however, and this is a simple (but unphysical) approximation that allows for a
straightforward handling of the kinematics in front of the continuum source. \\

\begin{figure}
\resizebox{8cm}{!}{\includegraphics[angle=0]{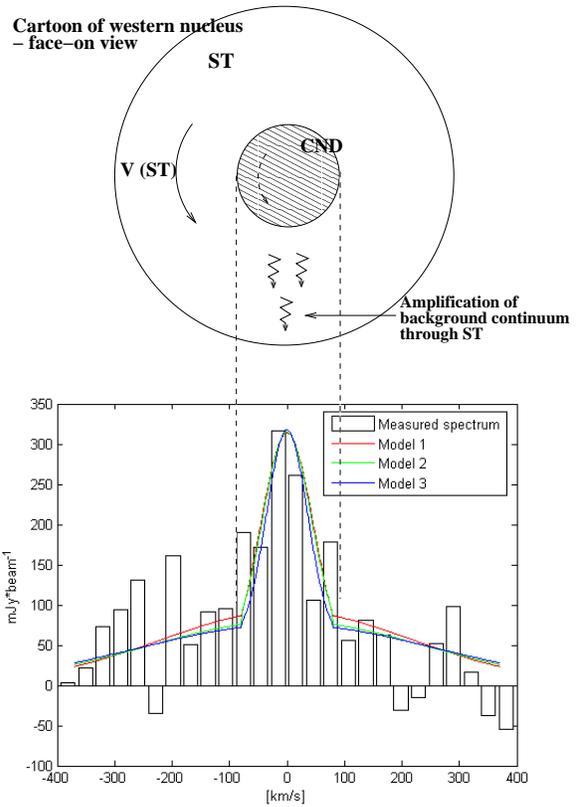}}
\caption{ \label{fig:fit} The figure shows a fit of a simple model (see text) to
the central spectrum of the western nucleus, and a face-on cartoon of the ST and CND.
The fit includes: 1) Emission from the surrounding 
torus: $T=T_{\rm x}(1-e^{-\tau})$; 2) Maser emission: $T=T_{\rm b}e^{-\tau}+T_{\rm x}(1-e^{-\tau})$, 
where $T_{\rm b}$ is (the enhanced) background temperature.
3) Emission from the CND: $T=T_0 e^{-\tau}$, where $T_0$ is the brightness temperature
from the inner disc (from a RADEX-model, with collisions only and modelled as a Gaussian,
emission damped by foreground cloud). 
There are three curves: {\bf Fit1}: $\tau=-$0.15, $T_0$=12, {\bf Fit2}: $\tau=-$0.08; $T_0$=30, 
and {\bf Fit3}: $\tau=-$0.04; $T_0$=40 
}
\end{figure}

In fig.~\ref{fig:fit} we show a model fit superposed on the nuclear spectrum towards
Arp~220-west. 
It consists of HNC 3--2 line emission contributions from the ST and its
amplification of the background continuum (the narrow feature) and the HNC emission 
from the CND. 
In our simplified assumption on the velocity (see above) the gas in front of the CND has
a velocity width of $\pm$80 \kms (similar to the fit of 60 \kms to the
observational data). Inside of this velocity range we get amplified
emission from the CND plus a contribution from the ST itself. Outside of this
velocity region there is only contribution from the ST and CND. We propose that the
populations are inverted throughout the ST and we have fixed the excitation temperature
in the ST to $-$40 K which gives 1\% population inversion. Other choices can be made
and we will discuss conditions for population inversion in the next section.

In the model fit the resulting average optical depth through the amplifying ST is
$\tau = -$0.16 (at $0.''4$ resolution ). There is also a contribution
to the HNC emission from the CND which is modelled to be about 4 K at line centre 
diluted in the 0.4\asec beam. The CND contributes all of the emission in the
line wings beyond a velocity of 300 \kms. 

We have assumed HNC column densities for the CND in
the range of $2 \times 10^{16} - 3 \times 10^{17}$ cm$^{-2}$ 
(i.e. an abundance relative to H$_2$ in the range $X$(HNC)=$2 \times 10^{-9} - 3\times 10^{-8}$),
and CND kinetic temperatures in the range 100 -- 170 K (close to the dust temperature). 
A high abundance in the CND will increase its line centre contribution and reduce
the resulting optical depth in the maser. (See discussion on HNC abundances in
Sect.~\ref{s:abundances}). If however the contribution from the CND becomes
too large it starts to affect the overall line shape, broadening it. Thus, for
the given conditions above, we find a range of maser optical depths of $\tau$=$-$0.04
to $\tau$=$-$0.25
where the higher absolute optical depth is for a small contribution from the CND.
Note, however, that the dusty graybody of the CND may prevent much line emission emerging
from the CND depending on dust optical depth, dust clumpiness and structure.

The shape of the observed spectrum can be reproduced within
the noise level. Testing a more sophisticated model requires higher signal-to-noise
observations at
higher resolution, where the ST and CND contributions can be resolved. Below we discuss
possible mechanisms for HNC population inversion.

\subsection{Conditions for population inversion}
\label{s:maser2}

In the previous section we propose that population inversion 
can take place between the rotational levels 3 and 2 and we show
the line profile in a simple case of amplification by the HNC molecule.
It is possible to either collisionally or radiatively pump HNC:

{\it a) Collisional pumping:} 
At gas densities approaching $n=10^6$ $\cmmd$
and kinetic temperatures around 150 K the HNC 3--2 line can become inverted. 
For a column density of $N$(HNC)=$10^{15}$ $\cmmt$ (assuming $N$(H$_2$)=$10^{24}$ through ST
and $X$(HNC)=$10^{-9}$), a line-width of 80 \kms, a continuum background
model as the one described in \citet{de07}, the HNC 3--2 line would be a maser with
$T_{\rm B}$=34 K, $\tau$=$-$0.16 and $T_{\rm ex}$=$-$58 K. This simulation
is done with RADEX code modified by us to include the mid-IR and IR transitions of HNC.
The on-line version of RADEX \citep{2007A&A...468..627V} does
currently not include these IR transitions of HNC.
This is to illustrate that HNC maser action can be achieved with the approximate
brightness, excitation and optical depth as indicated in the simple
model in the previous section. We furthermore note that in this model, the HNC 3--2
line is by far the most luminous of the rotational lines. The 1--0 and 2--1 lines
are also masering, but with lower brightness temperatures. The $J$=4--3 line is 
not masering and is fainter by a factor of 6.\\

{\it b) Radiative pumping:} 
Both HCN and HNC have degenerate bending modes in the IR through which they can
be pumped (see Sect.~\ref{s:intro}). 
HNC is pumped via 21.5 $\mu$m continuum and the pumping of HNC may start to become effective
when the IR background reaches a brightness temperature of $T_{\rm B} \approx$ 50 K at that
wavelength.
The inner 30 pc CND would then serve as a source of mid-IR photons to pump the HNC molecules in
the surrounding torus. According to \citet{de07}, the brightness temperature at 1.3 mm is 90 K.
\citet{sakamoto08} report the 860 $\mu$m brightness temperature of the west nucleus to be
(90-160) K, where the range corresponds to the range of possible source structure (i.e., a Gaussian or a disc). 
The nuclear mid-IR temperature may therefore be as high as 160 K, although
the mid-IR emission with brightness temperature of 160 K for the dust in the CND likely has a solid angle that is less
than 2$\pi$ as seen by the ST, it is luminous enough to affect significantly the excitation
of the ST - at least its inner parts. Furthermore, mid-IR imaging by
\citet{soifer99} showed that the brightness temperature is around
100 K in the $\approx$100 pc region around the western nucleus. The mid-IR photons for the
broad component may therefore be from the ~100 pc western disc itself. 
Modelling the geometry, radiation field, gas density and 
temperature will help determine what fraction of the HNC-emission from the ST that could be
affected by radiative pumping. 

\subsection{Physical conditions in the ST}
\label{s:st}

The broader part of the nuclear spectrum has a line width of 500 - 550 \kms
and a peak flux of 100-150 mJy (11 - 18 K)(fig.~\ref{fig:hnc32_spec}).
We identify this feature with the surrounding, 90 pc torus (ST) for which \citet{de07} find a
CO 2--1 brightness temperature of 40 K and \citet{sakamoto08} find a slightly higher CO 3--2 brightness
of 50 K. Thus, roughly, the CO 2--1/HNC 3--2 brightness temperature ratio
is only 2-4 for the ST outside of the luminous, narrow, maser-like feature.

Such a low line ratio is consistent with both possible pumping scenarios discussed
above. For the physical conditions suggested by \citet{de07} for
the ST it is difficult to get such a low CO/HNC line ratio. At $n=10^4$ 
$\cmmd$, the density is too low to effectively excite the HNC 3--2 line to
the required level unless the HNC abundances are extreme (see Sect.~\ref{s:abundances}
for a discussion of HNC abundances). However,
if the ST is very clumpy and clump densities can approach $n=10^6$ $\cmmd$ then
the observed line ratio can be understood. There is a large number of SNRs in this
area so it is likely that there is compressed, high-density gas around the SNRs.
Alternatively, the low CO/HNC line ratio can be a result of the HNC excitation 
being affected by mid-IR pumping since the mid-IR continuum would boost the HNC 3--2,
but not the CO emission.

\subsection{Alternative scenarios}
\label{s:alt}

Instead of originating as amplified emission in the ST, the narrow luminous HNC line may
emerge from the CND itself - but
then one would expect a much broader line width, similar to the suggested rotational
velocity \citep{de07} of 370 \kms. However, the line width of the inner CND is not well
constrained, so this is an option that warrants further study.

Another possibility is that the narrow feature emerges from a nuclear outflow. The shape
and width of the line would then be dependent on the viewing-angle of the disc, the outflow
velocity and the opening angle of the cone.

\subsection{The extended HNC 3--2 emission}
\label{s:ext}

Outside of the compact, and possibly amplified, emission of the western nucleus
we find HNC emission distributed on radii $0.''5 - 1''$ (150 - 350 pc) 
north and south of the western nucleus. This is not
an unreasonable scale in a merging system. Its north-south morphology
also resembles the bipolar morphology
of the OH maser emission from the western region of Arp~220
\citep{2003MNRAS.342..373R}. This is intriguing, since
both HNC and OH can undergo infrared radiative excitation.

\subsection{HNC abundances}
\label{s:abundances}

It is predicted in chemical steady state models and also by 
shock models that the HNC/HCN ratio drops with increasing temperature 
and gas density. This is supported by the fact 
that the measured HNC/HCN ratio is especially low in the vicinity of 
the hot core of Orion KL. 
This behaviour is to some degree caused by neutral-neutral chemistry in
warm ($T> 35$ K) dense environments: At high temperatures HNC
can be transferred into HCN via the reaction ${\rm HNC} + {\rm H} \to 
{\rm HCN} + {\rm H}$. Furthermore, HNC is destroyed through reactions with
atomic oxygen at high temperatures \citep[e.g.][]{schilke92,hirota98}.
Therefore, the HNC abundances in these regions are expected to be below $10^{-9}$
and factors of 10 -- 100 lower than those for HCN.

However, if the chemistry is dominated by ion-neutral processes,
the negative temperature dependence of the HNC
abundances is less pronounced or absent  even in warm regions,
which would be expected in a PDR or XDR.
In an XDR the ion-molecule chemistry is different from that of a PDR and may
lead to HNC abundances exceeding those of HCN even when temperatures are high
\citep{2005A&A...436..397M}.
For gas densities around $10^5$ $\cmmd$ in an XDR, $X$(HNC) may become twice
that of HCN. 

\subsubsection{The CND}

For the adopted densities and columns of the CND (see Sect.~\ref{s:maser1}), we would require
X-ray fluxes in excess of 10 erg cm$^{-2}$ s$^{-1}$ to affect the chemistry. 
One would have 100 erg cm$^{-2}$ s$^{-1}$ at 30 pc for an X-ray point source of
10$^{43}$ erg s$^{-1}$. For this, HNC abundances would be enhanced up to about 100 pc.
We find - using the models described in e.g. \citet{2005A&A...436..397M} -
HNC abundances increasing one order of magnitude with increasing 
X-ray flux (F$_{\rm X}$)(Tab~\ref{t:abundances}). Also the HCN abundances increase
with F$_{\rm X}$, but remain lower than $X$(HNC) with factors 1.2 to 1.8.

\begin{table}
\caption{\label{t:abundances} HNC and HCN abundances in the XDR-CND model$^a$}
\begin{tabular}{llc}
$X$(HNC) & $X$(HCN) & F$_{\rm X}$ (erg cm$^{-2}$ s$^{-1}$)\\
\hline
\hline \\ 
\pow{2.1}{-9} & \pow{1.7}{-9} &   \,\,\,\,\,\,1\\
\pow{8.8}{-9} & \pow{5.6}{-9} &  \,\,\,10 \\
\pow{2.9}{-8} & \pow{1.6}{-8} & 100 \\
\hline \\
\end{tabular} 

a) For a column of $N({\rm H}_2)=10^{25}$ $\cmmt$, density $n=10^5$ $\cmmd$
and temperature $T_{\rm k}$=150 K.
\end{table}

In Sect.~\ref{s:maser1} we discussed the contribution from the CND to the western
nuclear line profile and how it becomes merged with emission from the ST and the narrow,
amplified emission. 
At a density of $10^5$ $\cmmd$ and temperature 170 K even a large HNC abundance 
of $10^{-8}$ would not result in a very strong signal in the line wings (signal would be 2 K, or 17 mJy).
If the CND is clumpy with gas densities an order of magnitude higher than
the average density, the signal in the line wings could
start to be detectable even at the current noise level and resolution.
However, the CND is immersed in its own strong continuum emission which would affect the excitation
of HNC. A detailed model and higher signal-to-noise and resolution observations are 
required to more closely address the issue of HNC abundances in the core of Arp~220-west.

\subsubsection{The surrounding 90 pc torus, ST}

In the surrounding torus the H$_2$ column and average gas density are suggested
to be one order of magnitude lower than in the hot CND. 
As we mention in Sect.~\ref{s:st}, it would be difficult to explain the HNC line emission
of the ST from a medium of density of only $10^4$ $\cmmd$ assuming collisional excitation
unless HNC abundances were extremely high, $X$(HNC)=$10^{-6}$.
If we assume either radiative excitation or that the HNC emission is emerging
from a clumpy medium with higher densities than average, the bright HNC signal
is consistent with a range of abundances: $X$(HNC)=$10^{-9} - 10^{-8}$ which can
be fitted to both XDR and PDR models. 

\subsubsection{The extended HNC emission}

In Sect~\ref{s:ext} the possibility that part of the extended HNC emission is
associated with a nuclear outflow is mentioned. The bright HNC emission may be due
to significant HNC abundances in a dense medium and/or radiative excitation of HNC.
With no information on the density and temperature of the molecular medium here it
is difficult to estimate the HNC abundances.\\
The potentially high abundance of HNC in an
outflow is interesting since it has been viewed as a molecule
that becomes removed by shocks \citep[e.g.][]{schilke92}. 
\citet{loenen} have studied shock destruction of HNC in a star-bursting system. 
They find that a star formation rate of $\sim 10^2$ $M_\odot$ yr$^{-1}$ can bring down the HNC/HCN abundance
ratio by a factor of two in the nuclei of active galaxies.
In contrast, \citet{2004ApJ...612..342A} find that in some protostellar outflows
HNC abundances may become enhanced in the cavity walls.

We furthermore note the difference between the eastern and western nuclei.
The CO 3--2 luminosity difference between the two nuclei is less than two
\citep{sakamoto08}
while the HNC 3--2 luminosity of the eastern nucleus is only 15 - 20\% of that of
the western. If the HNC emission is 
enhanced by the IR fields and stimulated emission in the western nucleus, this demonstrates
the difference in degree and type of activity of the two nuclei. It is also possible that
the dense gas content in the eastern nucleus is lower than in the west.

\section{Outlook: Future observational tests}

There are a number of possible ways to further test the effects
of the intense continuum on the excitation of molecules in 
Arp~220. With better signal-to-noise and even higher resolution HNC data one should be able
to test hypotheses of radiative excitation, maser amplification and various
XDR/PDR models. Also the suggested difference in physical conditions
between the eastern and western nuclei should be further studied. 
It will be very useful to compare the HNC with high resolution observations of
other molecular lines, in particular other high density tracers. In an XDR scenario
a large abundance of CN is expected. Also CS, HCN, HCO$^+$, NO, HOC$^+$, SiO may be abundant
depending on X-ray flux, temperature, number- and column density of the XDR.
Rotational transitions of vibrationally excited HNC in the millimetre/submillimetre
are potentially detectable. 
Several of the suggested molecules
have important transitions in the infrared with large diagnostic potential.
The suggestion that the bipolar structure around the western nucleus is an outflow could be
addressed through observing HCO$^+$ -- which is a molecule that is often enhanced in
outflows \citep[e.g.][]{rawlings04}.


The fact that the HNC 3--2 emission from the western nucleus has its peak
intensity where CO 2--1 and 3--2 show
distinct absorption, is a strong indication that HNC may be a maser.
We can reproduce the HNC 3--2 spectrum with a simple maser model, but 
further tests are necessary. 
Higher resolution observations will reveal the presence of brighter maser-spots
if the emission is emerging from a 
clumpy medium. A weak, large-scale maser may however be difficult to distinguish
from strong thermal emission, even in a clumpy medium. Time variability
connected to the motions of maser spots and
variability in the nuclear source can also be investigated. High resolution
observations of other HNC transitions will help constrain radiative transport
models.

We furthermore note that the collisional excitation rates for HNC are not well known.
In models, it is therefore assumed that the rates are the same as for HCN.
Better collisional rates for HNC are necessary for fine-tuning of fits of observational
data to radiative transport and chemistry models.

\section{Conclusions}

    Luminous HNC 3--2 line emission is detected towards the western nucleus of the
    ULIRG Arp~220.
    About 50\% of the flux is emerging from the inner 180 pc consisting of an east-west
    rotating
    torus or disc surrounding a hot, dense, and dusty circum nuclear region of radius 30 pc.
    The emission
    consists of broad emission from the disc which we suggest is being pumped by mid-IR
    continuum from
    the buried circum nuclear region -- but also possibly by the surrounding torus itself. 
    Superposed on this emission is a narrow feature which may
    be the result of a weak HNC maser, enhancing the background continuum along the line of
    sight through
    the pumped disc. The emission feature coincides in position and velocity with absorption
    in the CO line.

    HNC 3--2 line emission is also found north and south of the nucleus. 
    This is either due to the presence of dense gas on scales of $0.''5 - 1''$ - and/or the
    mid-IR field is affecting the HNC excitation here as well. The spatial correlation with
    OH maser emission may suggest that part of the extended HNC emission is due to a nuclear outflow.

    The HNC emission is located mostly on the western side of Arp~220. Only faint emission
    is detected close to the eastern nucleus. This reflects a real difference in physical
    conditions between the two nuclei.

\begin{acknowledgements}
      We thank an anonymous referee for useful comments and suggestions that helped improve
the paper. The Submillimeter Array is a joint project between the Smithsonian
Astrophysical Observatory and the Academia Sinica Institute of Astronomy and
Astrophysics and is funded by the Smithsonian Institution and the Academia
Sinica. 
\end{acknowledgements}

\bibliographystyle{aa}
\bibliography{aalto_hnc_sma}

\begin{thebibliography}{21}
\expandafter\ifx\csname natexlab\endcsname\relax\def\natexlab#1{#1}\fi

\bibitem[{{Aalto} {et~al.}(2007){Aalto}, {Spaans}, {Wiedner}, \&
  {H{\"u}ttemeister}}]{aalto07a}
{Aalto}, S., {Spaans}, M., {Wiedner}, M.~C., \& {H{\"u}ttemeister}, S. 2007,
  \aap, 464, 193

\bibitem[{{Araya} {et~al.}(2004){Araya}, {Baan}, \& {Hofner}}]{araya04}
{Araya}, E., {Baan}, W.~A., \& {Hofner}, P. 2004, \apjs, 154, 541

\bibitem[{{Arce} \& {Sargent}(2004)}]{2004ApJ...612..342A}
{Arce}, H.~G. \& {Sargent}, A.~I. 2004, \apj, 612, 342

\bibitem[{{Cernicharo} {et~al.}(2006){Cernicharo}, {Pardo}, \&
  {Weiss}}]{cerni06}
{Cernicharo}, J., {Pardo}, J.~R., \& {Weiss}, A. 2006, \apjl, 646, L49

\bibitem[{{Downes} \& {Eckart}(2007)}]{de07}
{Downes}, D. \& {Eckart}, A. 2007, \aap, 468, L57

\bibitem[{{Hirota} {et~al.}(1998){Hirota}, {Yamamoto}, {Mikami}, \&
  {Ohishi}}]{hirota98}
{Hirota}, T., {Yamamoto}, S., {Mikami}, H., \& {Ohishi}, M. 1998, \apj, 503,
  717

\bibitem[{{Lepp} \& {Dalgarno}(1996)}]{1996A&A...306L..21L}
{Lepp}, S. \& {Dalgarno}, A. 1996, \aap, 306, L21

\bibitem[{{Loenen} {et~al.}(2008){Loenen}, {Spaans}, {Baan}, \&
  {Meijerink}}]{loenen}
{Loenen}, A.~F., {Spaans}, M., {Baan}, W.~A., \& {Meijerink}, R. 2008, \aap,
  488, L5

\bibitem[{{Maloney} {et~al.}(1996){Maloney}, {Hollenbach}, \&
  {Tielens}}]{1996ApJ...466..561M}
{Maloney}, P.~R., {Hollenbach}, D.~J., \& {Tielens}, A.~G.~G.~M. 1996, \apj,
  466, 561

\bibitem[{{Meijerink} \& {Spaans}(2005)}]{2005A&A...436..397M}
{Meijerink}, R. \& {Spaans}, M. 2005, \aap, 436, 397

\bibitem[{{Meijerink} {et~al.}(2006){Meijerink}, {Spaans}, \&
  {Israel}}]{2006ApJ...650L.103M}
{Meijerink}, R., {Spaans}, M., \& {Israel}, F.~P. 2006, \apjl, 650, L103

\bibitem[{{Meijerink} {et~al.}(2007){Meijerink}, {Spaans}, \&
  {Israel}}]{2007A&A...461..793M}
{Meijerink}, R., {Spaans}, M., \& {Israel}, F.~P. 2007, \aap, 461, 793

\bibitem[{{Rawlings} {et~al.}(2004){Rawlings}, {Redman}, {Keto}, \&
  {Williams}}]{rawlings04}
{Rawlings}, J.~M.~C., {Redman}, M.~P., {Keto}, E., \& {Williams}, D.~A. 2004,
  \mnras, 351, 1054

\bibitem[{{Rovilos} {et~al.}(2003){Rovilos}, {Diamond}, {Lonsdale}, {Lonsdale},
  \& {Smith}}]{2003MNRAS.342..373R}
{Rovilos}, E., {Diamond}, P.~J., {Lonsdale}, C.~J., {Lonsdale}, C.~J., \&
  {Smith}, H.~E. 2003, \mnras, 342, 373

\bibitem[{{Sakamoto} {et~al.}(1999){Sakamoto}, {Scoville}, {Yun}, {Crosas},
  {Genzel}, \& {Tacconi}}]{sakamoto99}
{Sakamoto}, K., {Scoville}, N.~Z., {Yun}, M.~S., {et~al.} 1999, \apj, 514, 68

\bibitem[{{Sakamoto} {et~al.}(2008){Sakamoto}, {Wang}, {Wiedner}, {Wang},
  {Peck}, {Zhang}, {Petitpas}, {Ho}, \& {Wilner}}]{sakamoto08}
{Sakamoto}, K., {Wang}, J., {Wiedner}, M.~C., {et~al.} 2008, \apj, 684, 957

\bibitem[{{Salter} {et~al.}(2008){Salter}, {Ghosh}, {Catinella}, {Lebron},
  {Lerner}, {Minchin}, \& {Momjian}}]{salter08}
{Salter}, C.~J., {Ghosh}, T., {Catinella}, B., {et~al.} 2008, \aj, 136, 389

\bibitem[{{Schilke} {et~al.}(1992){Schilke}, {Walmsley}, {Pineau des F\^orets},
  {Roueff}, {Flower}, \& {Guilloteau}}]{schilke92}
{Schilke}, P., {Walmsley}, C.~M., {Pineau des F\^orets}, G., {et~al.} 1992,
  \aap, 256, 595

\bibitem[{{Soifer} {et~al.}(1999){Soifer}, {Neugebauer}, {Matthews}, {Becklin},
  {Ressler}, {Werner}, {Weinberger}, \& {Egami}}]{soifer99}
{Soifer}, B.~T., {Neugebauer}, G., {Matthews}, K., {et~al.} 1999, \apj, 513,
  207

\bibitem[{{Tielens} \& {Hollenbach}(1985)}]{1985ApJ...291..722T}
{Tielens}, A.~G.~G.~M. \& {Hollenbach}, D. 1985, \apj, 291, 722

\bibitem[{{van der Tak} {et~al.}(2007){van der Tak}, {Black}, {Sch{\"o}ier},
  {Jansen}, \& {van Dishoeck}}]{2007A&A...468..627V}
{van der Tak}, F.~F.~S., {Black}, J.~H., {Sch{\"o}ier}, F.~L., {Jansen}, D.~J.,
  \& {van Dishoeck}, E.~F. 2007, \aap, 468, 627

\end{thebibliography}

\end{document}